\documentclass[applsci,article,accept]{Definitions/mdpi} 
\firstpage{1} 
\pubvolume{1}
\issuenum{1}
\articlenumber{0}
\pubyear{2025}
\copyrightyear{2025}
\datereceived{ } 
\daterevised{ } 
\dateaccepted{ } 
\datepublished{ } 
\hreflink{https://doi.org/} 

\Title{Enhancing Retrieval-Augmented Generation with Entity Linking for Educational Platforms}

\TitleCitation{Enhancing Retrieval- Augmented Generation with Entity Linking for Educational Platforms}

\Author{Francesco Granata $^{1}$\orcidA{}, Francesco Poggi $^{2}$\orcidC{}, Misael Mongiovì $^{1,2}$\orcidB{}}

\AuthorNames{Francesco Granata, Misael Mongiovì and Francesco Poggi}

\isAPAStyle{%
       \AuthorCitation{Lastname, F., Lastname, F., \& Lastname, F.}
         }{%
        \isChicagoStyle{%
        \AuthorCitation{Lastname, Firstname, Firstname Lastname, and Firstname Lastname.}
        }{
        \AuthorCitation{Granata, F.; Mongiovì, M.; Poggi, F.}
        }
}

\address{%
$^{1}$ \quad Department of Mathematics and Computer Science, University of Catania, Italy; grnfnc03t17c351z@studium.unict.it, misael.mongiovi@unict.it\\
$^{2}$ \quad Institute of Cognitive Sciences and Technologies (ISTC), National Research Council of Italy (CNR); francesco.poggi@cnr.it\\
}

\corres{Correspondence: francesco.poggi@cnr.it; Tel.: +39 0512088202
}

\abstract{In the era of Large Language Models (LLMs), Retrieval-Augmented Generation (RAG) architectures are gaining significant attention for their ability to ground language generation in reliable knowledge sources. Despite their impressive effectiveness in many areas, RAG systems based solely on semantic similarity often fail to ensure factual accuracy in specialized domains, where terminological ambiguity can affect retrieval relevance. This study proposes an enhanced RAG architecture that integrates a factual signal derived from Entity Linking to improve the accuracy of educational question-answering systems in Italian. 
The system includes a Wikidata-based Entity Linking module and implements three re-ranking strategies to combine semantic and entity-based information: a hybrid score weighting model, reciprocal rank fusion, and a cross-encoder re-ranker. Experiments were conducted on two benchmarks: a custom academic dataset and the standard SQuAD-it dataset. Results show that, in domain-specific contexts, the hybrid schema based on reciprocal rank fusion significantly outperforms both the baseline and the cross-encoder approach, while the cross-encoder achieves the best results on the general-domain dataset. These findings confirm the presence of an effect of domain mismatch and highlight the importance of domain adaptation and hybrid ranking strategies to enhance factual precision and reliability in retrieval-augmented generation. They also demonstrate the potential of entity-aware RAG systems in educational environments, fostering adaptive and reliable AI-based tutoring tools.}


\begin{document}

\section{Introduction}


In the current technological landscape, dominated by \textit{Large Language Models (LLMs)}, the way we approach information and data has profoundly changed. 
These models are capable of producing fluent and context-aware text across a wide range of domains, making them powerful tools for knowledge access, task automation, and decision support. 
Their impressive generative capabilities, however, come with inherent limitations.
Despite these strengths, LLMs are not always the best way to access knowledge, especially in critical, specialized or high-precision domains, because they can sometimes provide incorrect, inconsistent or unverifiable information. 
This phenomenon - commonly referred to as \textit{hallucination} \cite{Huang2025} - poses significant risks when accuracy and reliability are essential, such as in scientific research ~\cite{Zhang_2025}, medicine~\cite{Asgari2025,Pal_2023,haq2026heal}, and education~\cite{Qian2026,Vrdoljak2025}. 

To address this problem and increase factual consistency, \textit{Retrieval-Augmented Generation (RAG)} architectures have proven to be one of the most effective approaches. 
By combining the generative abilities of LLMs with external knowledge sources, RAG systems can retrieve relevant documents or passages and ground the model’s output in real, verifiable information. This allows them to provide precise and up-to-date answers without requiring model retraining or large-scale parameter updates. However, these systems are not perfect: most retrieval components rely heavily on semantic similarity, which can fail to adequately capture subtle distinctions in technical language, polysemous terms, or domain-specific terminology. 
As a result, retrieval errors propagate into the generation phase, reducing the overall reliability of the system. 

These limitations become particularly evident in the educational domain, where materials often contain highly specialized vocabulary, hierarchical conceptual structures, and terminology that differs across subjects or instructional levels. 
Standard RAG approaches may struggle to identify the correct referents for ambiguous or overlapping terms, leading to suboptimal retrieval and factually incorrect responses. 
To improve performance in this context we explore the integration between \textit{Entity Linking} and RAG architectures.
Entity Linking provides a structured mechanism for identifying and disambiguating mentions of concepts within text, mapping them to unique identifiers in a knowledge base. This additional layer of semantic grounding has the potential to refine retrieval queries, reduce ambiguity, and enhance the factual stability of generated outputs.
The aim of this work is to investigate how Entity Linking can enhance retrieval accuracy and factual reliability in RAG systems applied to educational content.

\subsection{Related Work}

To improve the factual grounding of large language models (LLMs), several studies have examined \textit{Retrieval-Augmented Generation} (RAG) systems as a promising framework to address this open research challenge. The original architecture proposed by Lewis et al. (2020)~\citep{lewis2020retrieval} integrates a neural retriever with a generative model, enabling language models to access external knowledge sources dynamically. A standard RAG pipeline can be conceptually divided into two main components: a retrieval phase and a generation phase. 

Historically, the retrieval component has relied on sparse, lexical matching techniques, with BM25~\citep{robertson2009probabilistic} standing as the most widely adopted probabilistic model based on exact keyword overlap. While highly effective for specific terminology, sparse methods struggle with the vocabulary mismatch problem. Consequently, recent RAG architectures have heavily shifted towards dense retrieval, utilizing neural embeddings based on semantic similarity to capture deep relationships.
Alternative retrieval approaches have also been explored. For example, Mongiovì et al. (2024)~\citep{mongiovi2024graal} introduced \textit{GRAAL}, a graph-based retrieval system that collects related passages across multiple documents by exploiting inter-entity relationships.

Despite these advancements, many dense retrieval models still face challenges in capturing fine-grained semantic meaning, particularly in specialized domains that contain ambiguous or domain-specific terminology~\cite{Ji2023}. Because dense models often underperform on exact string matching for rare entities or acronyms, state-of-the-art approaches increasingly employ hybrid search strategies~\cite{weinberg2026hybrid}. These mixed methods combine the exact lexical precision of sparse models (like BM25) with the semantic generalization of dense embeddings, typically merging the distinct retrieval lists through aggregation algorithms such as Reciprocal Rank Fusion (RRF)~\citep{rrf}.

Our work extends this hybrid retrieval paradigm by replacing the purely lexical sparse retriever with an explicit, knowledge-grounded signal. Specifically, \textit{Entity Linking} (EL) has been proposed as a method to improve the disambiguation of named entities by aligning textual mentions with entries in structured knowledge bases such as Wikidata \citep{wikidata2025} or DBpedia. Shen et al. (2015)~\citep{shen2014entity} provided an extensive overview of EL challenges and techniques. Möller et al. (2022)~\citep{moller2022survey} presented a comprehensive survey of Wikidata-based EL methods and datasets. Several frameworks have been developed in this area, including BLINK~\citep{wu2019scalable}, ReLiK~\citep{orlando2024relik}, and OpenTapioca~\citep{delpeuch2019opentapioca}, each adopting different architectures and disambiguation strategies. Most current EL systems, however, have been primarily developed for the English language, which limits their direct applicability in multilingual or non-English domains.

At the same time, \textit{Educational Artificial Intelligence (EAI)} has become a rapidly expanding field of research. AI-based systems are increasingly applied to support learning, personalize content delivery, and facilitate knowledge acquisition~\cite{Furini202298}. Mageira et al. (2022)~\citep{mageira2022educational} discussed the pedagogical potential of educational chatbots designed for content and language-integrated learning. More recently, Swacha et al.(2025)~\citep{swacha2025retrieval} and  Li et al. (2025)~\citep{li2025retrieval} provided a systematic survey of RAG applications in education, illustrating how retrieval-augmented approaches can enhance information access and automated question answering in academic contexts.
More broadly, recent studies have explored the integration of LLM-based architectures within domain-specific decision-support systems, including frameworks that combine language models with multimodal data processing and explainable reasoning for applications such as healthcare decision making~\cite{HAQ2026102696}.

The present work builds upon these lines of research by combining RAG and Entity Linking within a hybrid retrieval architecture applied to the educational domain in Italian. While prior research, such as Shlyk et al. (2024)~\citep{shlyk2024real}, has explored the integration of RAG and EL for biomedical concept recognition, our approach differs in both pipeline design and linguistic scope, focusing on hybrid retrieval in an Italian-language educational setting.

\subsection{Research Objectives and Contributions}

The objective of this study is to investigate how the integration of an \textit{Entity Linking} component can enhance the performance of \textit{Retrieval-Augmented Generation} (RAG) systems in specialized educational contexts. In particular, the research focuses on developing and evaluating a hybrid retrieval architecture that combines semantic similarity and entity-based information to improve the relevance and factual reliability of retrieved passages. The study is conducted entirely on Italian-language educational data, addressing the challenge of adapting retrieval and generation techniques to a non-English education domain.

The main contributions of this work can be summarized as follows:
\begin{enumerate}
    \item We design and implement \textit{ELERAG}: a hybrid RAG architecture that integrates a Wikidata-based Entity Linking module to incorporate entity-level knowledge during retrieval.
    \item We evaluate the efficacy of our proposed \textit{RRF-Based Re-ranking} strategy by comparing it against a baseline Weighted-Score Re-ranking, a high-complexity RRF + Cross-Encoder Re-ranking and a standalone Cross-Encoder Re-ranking.
    \item We analyze the results across both custom educational data and standard benchmarks, highlighting the effects of domain adaptation and linguistic specificity on retrieval performance.
    \item We provide clear experimental evidence of a \textit{domain mismatch}, demonstrating that a domain-adapted hybrid model can outperform a generic State-Of-The-Art re-ranker on specialized data, whereas the SOTA model excels on standard benchmarks.
\end{enumerate}

By addressing the intersection of RAG, Entity Linking, and Educational AI, this study contributes to ongoing efforts to improve factual grounding and domain-sensitive retrieval in large language model applications. It introduces and analyzes a new class of hybrid architectures designed to support more trustworthy, transparent and pedagogically aligned AI-driven educational tools.

This paper is organized as follows: Section \ref{sec:methodology} details the methodology, describing the hybrid RAG architecture, the Entity Linking module, and the implemented re-ranking strategies. Section \ref{sec:experimentalsetup} presents the experimental setup, including the construction of the custom educational dataset, the benchmarks, and the evaluation metrics. Section \ref{sec:results} reports the quantitative and qualitative results obtained from the experiments. Finally, Section \ref{sec:discussion} discusses the findings and draws the main conclusions, highlighting the domain mismatch phenomenon and outlining limitations and future directions.


\section{Methodology}\label{sec:methodology}

This section details the architecture and implementation of \textit{ELERAG} (Entity Linking Enhanced RAG), the hybrid retrieval system proposed in this study. We first describe the baseline RAG configuration used as a reference, followed by the Entity Linking module designed to improve semantic disambiguation. Next, we present the core re-ranking strategy adopted in \textit{ELERAG}, along with alternative strategies implemented for comparative analysis. Finally, we illustrate the complete end-to-end workflow of the proposed method.

\subsection{RAG baseline}\label{subsec:rag_baseline}

We built a baseline \textit{Retrieval-Augmented Generation} (RAG) system as a starting point for our architecture.
This baseline follows the standard structure of embedding-based retrieval combined with generative models, a paradigm introduced by Lewis et al (2020) \citep{lewis2020retrieval} and widely explored in recent educational applications~\citep{swacha2025retrieval, li2025retrieval}.
Our approach combines the vector store \textit{FAISS} \citep{johnson2019faiss} with \texttt{multilingual-e5-large} \citep{wang2024multilingual} as an embedding model and GPT-4o \citep{openai2023gpt4o} as a generator.

Each chunk in the corpus was encoded into a fixed-length 1024-dimensional vector representation using the \texttt{multilingual-e5-large} model by \textit{Sentence Transformers} \footnote{\url{https://sbert.net}}, chosen for its strong multilingual retrieval capabilities and competitive performance on cross-lingual benchmarks.
The resulting embeddings were normalized and stored in a \textit{FAISS} index employing an inner-product \footnote{for normalized vectors it is equivalent to cosine similarity} similarity metric to enable efficient dense retrieval at query time.
During inference, a user query is first encoded with the same embedding model as the corpus; then, the top-$K$ most similar vectors are retrieved, and the corresponding document chunks are returned. 
These retrieved passages are concatenated to form a context window, which is then provided as additional input to the generative model. 

We used \textit{GPT-4o} as the generator, prompting it to answer based solely on the retrieved content and to abstain when sufficient evidence is not present in the context. Crucially, the prompt explicitly instructs the model to cite the source chunk IDs for every piece of information used in the answer. Consequently, the generation phase acts as a final refinement stage: only the chunks actually cited in the generated response are considered ``retrieved'' by the full system, while uncited chunks—even if present in the context window—are effectively discarded as irrelevant.
This configuration serves as the reference setup for evaluating the contribution of entity-level enrichment and alternative ranking strategies in the following sections.

\subsection{Entity Linking Module}\label{subsec:ELmodule}

While the baseline RAG system relies solely on semantic similarity for retrieval, this approach often struggles with domain-specific ambiguity and the presence of polysemous terms. 
In educational material, where concepts may appear with slight linguistic variations across disciplines, pure embedding similarity can retrieve semantically close but contextually irrelevant chunks. 
To address this limitation, we integrated an \textit{Entity Linking} (EL) module designed to ground text spans to canonical entities within \textit{Wikidata}~\citep{wikidata2025}.

Every chunk in the dataset was pre-processed to extract named entities using the \textit{SpaCy} pipeline~\citep{spacy2023}. Specifically, we employed the \texttt{it\_core\_news\_lg} \footnote{\url{https://spacy.io/models/it\#it_core_news_lg}} model, a large pre-trained pipeline for the Italian language trained on a massive corpus of news and media text, chosen for its superior accuracy in Named Entity Recognition (NER) compared to smaller variants. 
Since the corpus originates from automatically transcribed lecture speech, entity mentions may occasionally appear in abbreviated forms or lexical variants. The linking step partially mitigates this issue by relying on the Wikidata knowledge base, which maintains extensive lists of aliases and alternative labels for each entity. As a result, common variations (e.g., surnames, acronyms, or shortened forms) can still be resolved to the correct Wikidata identifier during candidate retrieval. 
In addition, the upstream transcription stage uses the Whisper Turbo ASR model, which provides high accuracy on proper names in lecture-style speech. This reduces the frequency of phonetic transcription errors that could otherwise affect the subsequent NER and linking stages.

Before developing this custom solution, we experimented with standard state-of-the-art Entity Linking systems such as \textit{BLINK}~\citep{wu2020blink}. However, since these models are primarily optimized for English, they yielded unsatisfactory results when applied to our Italian educational corpus. Consequently, we opted for a lightweight, API-based approach tailored to our specific language requirements.
For each detected entity mention, a candidate list of Wikidata entities was retrieved through the public Wikidata API.

To select the best candidate, a hybrid scoring function was developed combining two signals:

\begin{enumerate}
    \item \textit{Popularity} — computed as the inverse of the candidate rank in the list returned by Wikidata:
    \begin{equation}
        \text{popularity} = \frac{1}{\text{rank} + 1}
    \end{equation}

    \item \textit{Semantic similarity} — obtained using \texttt{multilingual-e5-large}, applied to the mention context (the sentence where the entity was detected) and the concatenation of the candidate’s \textit{label} and its \textit{description}.
\end{enumerate}

The final score is computed as:
\begin{equation}
    \text{HybridScore} = \alpha \cdot \text{similarity} + (1 - \alpha) \cdot \text{popularity}
\end{equation}

where $\alpha$ ($= 0.9$) controls the balance between semantic and popularity signals. Note that the similarity score considers the whole mention context, therefore it is able to distinguish among multiple entities which correspond to the same mention.

The candidate with the highest score is selected as the final linked entity and stored, together with all metadata and supporting information, in a JSON structure with the addition of linked entities.

Consequently, if an entity mention is not detected due to transcription noise or spelling variations, the entity-based signal is simply absent. In such cases, the system naturally falls back to the dense semantic retrieval component, ensuring that the overall retrieval process remains robust.

\subsection{Ranking Strategies and Integration into the Retrieval Pipeline}

The enrichment introduced by the \textit{Entity Linking} module was used to enhance the retrieval phase by re-ranking the initially retrieved chunks from the FAISS index. 
Specifically, after dense retrieval, an additional entity-aware re-ranking stage was applied. 

\paragraph{Proposed Strategy: RRF-Based Re-ranking.}
The core ranking strategy adopted in our \textit{ELERAG} architecture is based on \textit{Entity-Aware Reciprocal Rank Fusion (RRF)}.
In this configuration, chunks are independently ranked according to their dense score (semantic similarity) and their entity score (factual overlap). The two distinct rankings are then fused using the RRF algorithm \citep{rrf}, which assigns a joint score based on the rank position in each list:
\begin{equation}
    \text{score}_{\text{RRF}} = \frac{1}{K + \text{rank}_{\text{dense}}} + \frac{1}{K + \text{rank}_{\text{entity}}}
\end{equation}
where $K=60$\footnote{60 is the value suggested by the authors of RRF and is a de facto standard \citep{rrf,tang2024found}} is the standard smoothing constant. This approach was selected as the primary method for \textit{ELERAG} because it robustly balances semantic relevance with factual entity matching without requiring the manual tuning of weights or the high computational cost of cross-encoders. As demonstrated in Section~\ref{sec:results}, this strategy yielded the best performance on our specialized educational dataset.

An intrinsic architectural advantage of this specific RRF implementation is its graceful degradation. In edge cases where the user query contains no recognizable named entities, the entity matching score evaluates to zero for all candidate chunks. Consequently, the entity-based sorting step preserves the initial order, meaning the entity rank ($r_{EL}$) simply mirrors the dense rank ($r_{dense}$). The final RRF calculation smoothly proceeds without altering the relative ordering, allowing the system to seamlessly fall back to the Standard Dense RAG baseline without computational interruptions or penalty assignments.

\paragraph{Comparative Strategies.}
To rigorously evaluate the effectiveness of our proposed strategy, we implemented three alternative re-ranking methods for comparison:

\begin{enumerate}
    \item \textit{Hybrid Sparse-Dense Baseline.}
    To thoroughly evaluate the contribution of the Entity Linking module against traditional exact-match approaches, we implemented a strong hybrid search baseline. This strategy combines the semantic representations of the \texttt{multilingual-e5-large} dense retriever with the exact keyword matching of a standard BM25 \citep{robertson2009probabilistic} sparse retriever (via the \texttt{rank\_bm25} library). To ensure a fair comparison with ELERAG, the sparse and dense signals were fused using the identical Reciprocal Rank Fusion (RRF) mechanism employed in our proposed pipeline.

    \item \textit{Weighted-Score Re-ranking.} 
    Each chunk was evaluated using a combined score:
    \begin{equation}
        \text{final\_score} = \text{dense\_score} + \beta \cdot \text{entity\_score}
    \end{equation}
    where \text{dense\_score} represents the cosine similarity between the query and the chunk embedding computed by \texttt{multilingual-e5-large}, 
    and the \text{entity\_score} represents the recall-oriented overlap between the set of query linked entities ($Q_E$) and the set of chunk linked entities ($C_E$): 
    \begin{equation}
        \text{entity\_score} = \frac{|Q_E \cap C_E|}{|Q_E|} \text{ if } |Q_E| > 0 \text{ else } 0
    \end{equation}
    The hyperparameter $\beta$ controls the relative contribution of entity-based evidence compared to semantic similarity.

    \item \textit{RRF + Cross-Encoder Re-ranking}. 
    This configuration explores a multi-stage refinement pipeline. After retrieving an initial pool of candidates via dense search and fusing them through RRF, the highest-ranking chunks are re-scored by a transformer-based Cross-Encoder~\citep{crossencoder} (namely, \texttt{mmarco-mMiniLMv2-L12-H384-v1}). Unlike bi-encoders (like E5), the Cross-Encoder processes the query and document simultaneously, capturing finer semantic nuances at a higher computational cost. This setup tests whether adding a deep semantic layer \textit{on top} of the entity signal yields further improvements.

    \item \textit{Standalone Cross-Encoder}. 
    To isolate the effect of the neural re-ranker and serve as a strict baseline for State-of-the-Art semantic retrieval without entity enhancement, we implemented a pure Cross-Encoder pipeline. The candidates retrieved directly by the dense vector index are re-ranked by the Cross-Encoder model, bypassing any entity-based signal or RRF fusion. The top candidates are then passed to the generator. This configuration allows us to verify whether the domain-specific entity signal adds distinct value compared to a generic, high-performance deep learning re-ranker.
\end{enumerate}

In all configurations, the progressive re-ranking across stages reduces the candidate set size, 
allowing the system to focus on the most promising passages for the subsequent generation phase. 

\begin{figure}[H]
    \centering
    \includegraphics[width=\textwidth]{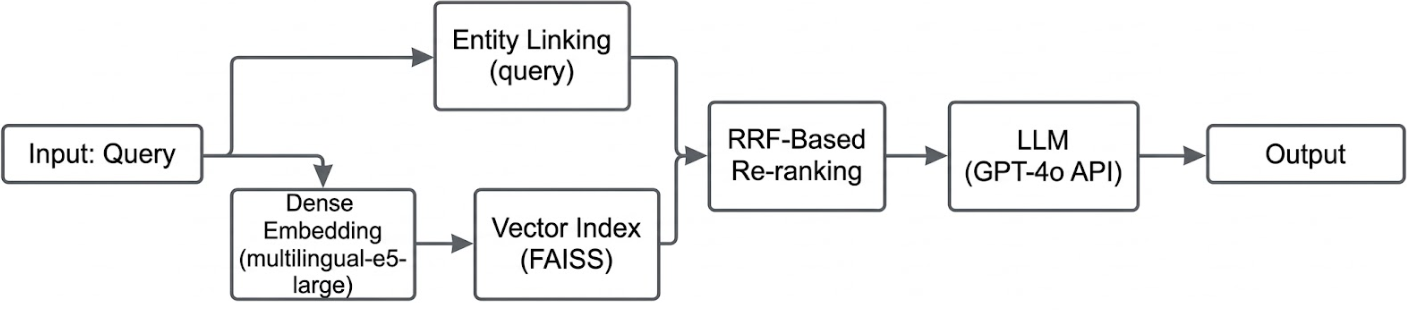}
    \caption{Architectural schema of the proposed \textit{ELERAG} method. The system integrates parallel retrieval paths—semantic dense retrieval and entity linking—fusing them via an RRF-based re-ranking module to ground the LLM generation.}
    \label{fig:elerag_schema}
\end{figure}

Figure~\ref{fig:elerag_schema} illustrates the complete workflow of our proposed \textit{ELERAG} method. The architecture processes the query in parallel streams—extracting entity-based features and computing dense embeddings—before fusing the results via the RRF module to feed the LLM.
The other experimental configurations can be understood as variations of this schema: the \textit{Standard RAG} baseline utilizes only the lower branch (Dense Embedding $\rightarrow$ Vector Index $\rightarrow$ LLM), bypassing the Entity Linking and Re-ranking stages. The \textit{Weighted-Score Re-ranking} configuration replaces the RRF block with the linear combination strategy described above. The \textit{RRF + Cross-Encoder Re-ranking} configuration adds a further refinement block between the RRF stage and the LLM generation. Finally, the \textit{Standalone Cross-Encoder} utilizes only the lower branch (Dense Embedding $\rightarrow$ Vector Index $\rightarrow$ Cross-Encoder $\rightarrow$ LLM) bypassing the Entity Linking and Re-ranking stages and adding the Cross-Encoder step.

\section{Experimental Setup}\label{sec:experimentalsetup}

This section presents the experimental setup designed to assess the effectiveness of the proposed \textit{ELERAG} architecture. 
It describes the datasets used for evaluation, both custom and standard, and the criteria adopted to ensure comparability across different domains. 
Then, it details the metrics employed to quantify retrieval and generation performance, and the evaluation methods applied to each system configuration. 
Together, these components define a consistent framework for analyzing and comparing the proposed hybrid retrieval models.

\subsection{Educational Data}\label{sec:dataset}

The dataset used in this study was constructed from two Italian university courses, namely \textit{Applied Economics} and \textit{Language and Communication}, offered by an Italian telematic university. 
Italian was selected as the target language because it allowed the construction of a novel evaluation dataset derived from educational lecture material, which differs significantly from widely used public QA benchmarks. From a resource perspective, Italian also represents an intermediate scenario: while supported by Wikipedia and Wikidata, its coverage is substantially smaller than English, making it suitable for evaluating the robustness of the proposed approach outside extremely resource-rich settings.
Each course consisted of several video lectures, totalling 50 classes over 32 hours of material, which were first transcribed into text using the \textit{Whisper Turbo} automatic speech recognition model~\citep{radford2022whisper}. 
The transcription process also generated precise word-level timestamps, which were later used to associate each text segment with its corresponding temporal interval in the original lecture.

To prepare the textual material for retrieval, the transcribed lectures were segmented into coherent chunks using the SpaCy Python library~\citep{spacy2023}.
To ensure semantic coherence, segmentation was strictly constrained to sentence boundaries: sentences were grouped to form chunks with a length between 20 and 300 tokens.
This strategy prevents cutting sentences in the middle, preserving the syntactic and semantic integrity of the text. Crucially, because the chunking process respects sentence boundaries, the local context required for Named Entity Recognition remains fully intact. Consequently, when the stored chunks are subsequently processed by the SpaCy NER pipeline (prior to the Entity Linking module described in Section \ref{subsec:ELmodule}), the quality of entity extraction is not negatively affected by arbitrary word-level cuts.

To quantify the density of factual and domain-specific information within the corpus, we computed the entity coverage metrics following the offline Entity Linking phase. Out of a total of 676 chunks, 540 chunks contain at least one successfully linked Wikidata entity, yielding an overall chunk coverage of 79.88\%. In total, 4,393 entities were extracted and disambiguated across the corpus, averaging 6.50 linked entities per chunk. This high entity density confirms that the educational narrative is rich in specific terminology, concepts, and factual references, providing a solid foundation for the subsequent entity-based re-ranking phase.

\subsection{Evaluation setup}

The data produced by the procedure in Section~\ref{sec:dataset} does not contain question-answer pairs, necessary to evaluate our approach. 
Therefore, we adopted a dual evaluation strategy that combines a custom benchmark, specifically generated from the course material using the GPT-4o API, and a standard benchmark (SQuAD-it~\citep{croce2018squadit}). 
This setup allows us to assess the system’s performance both within a specialized educational context and in a general-domain question answering scenario.

The custom benchmark was automatically constructed from the lecture corpus discussed in Section~\ref{sec:dataset}.
Using the GPT-4o API, we prompted the model to generate three types of questions---\textit{factual}, \textit{synthesis}, and \textit{inference}---together with their corresponding gold answers and relevant document references.
The resulting benchmark contains 69 questions, each represented as a structured record including query, question type, gold answer, relevant documents, and additional metadata.

After generation, the dataset underwent a two-step validation process. First, a secondary GPT-4o prompt was used to verify the correctness and consistency between questions and their corresponding answers, automatically filtering invalid records. Subsequently, a manual curation was performed collaboratively by the authors, acting as domain experts, to identify and filter out any residual ambiguous, ill-formed, or hallucinated questions that might have bypassed the automated check. Because this final refinement was conducted as an expert consensus process rather than a parallel labeling task by independent annotators, standard inter-annotator agreement metrics were not applicable. All authors reviewed the generated questions and reached full consensus on the final selection. This rigorous two-step process ensured the high quality of the final set of 69 questions used for evaluation.

\paragraph{Standard Benchmark (SQuAD-it).}
For comparison with a standard task, we employed the SQuAD-it benchmark, a widely used dataset for Italian question answering derived from Wikipedia.
The dataset consists of a collection of passages (contexts), and for each passage, a set of question-answer pairs. To adapt it to our retrieval evaluation setting, we treated each Wikipedia passage as a distinct "chunk" and indexed them exactly as we did for the educational dataset.
In this configuration, for each query, the passage referenced in the original dataset (the \textit{context}) is treated as the unique \textit{gold answer} to be retrieved.

\subsection{Evaluation Metrics}

The following metrics were used to compare the retrieval and generation performance of the different systems:

\begin{itemize}
    \item \textit{Exact Match (EM):} Proportion of queries for which the first retrieved document exactly matches the gold answer.
    \[
    \text{EM} = \frac{1}{N} \sum_{i=1}^{N} \mathbf{1}\{\text{retrieved\_docs}_i[0] = \text{gold\_answer}_i\}
    \]

    \item \textit{Recall@k:} Proportion of relevant documents that appear among the top-$k$ retrieved results.
    \[
    \text{Recall@k} = \frac{1}{N} \sum_{i=1}^{N} \frac{|\{doc \in \text{retrieved\_docs}_i[:k] : doc \in \text{relevant\_docs}_i\}|}{|\text{relevant\_docs}_i|}
    \]

    \item \textit{Precision@k:} Proportion of the top-$k$ retrieved documents that are relevant.
    \[
    \text{Precision@k} = \frac{1}{N} \sum_{i=1}^{N} \frac{|\{doc \in \text{retrieved\_docs}_i[:k] : doc \in \text{relevant\_docs}_i\}|}{k}
    \]

    \item \textit{Mean Reciprocal Rank (MRR):} Mean reciprocal rank of the first occurrence of a relevant document. Two distinct versions were computed:
    \begin{itemize}
        \item MRR based on the \texttt{gold\_answer}:
        \[
        \text{MRR\_gold} = \frac{1}{N} \sum_{i=1}^{N} \frac{1}{\text{rank}_i(\text{gold\_answer}_i)}
        \]
        \item MRR based on all relevant documents:
        \[
        \text{MRR\_rel\_docs} = \frac{1}{N} \sum_{i=1}^{N} \frac{1}{\text{rank}_i(\text{first\_relevant\_doc})}
        \]
    \end{itemize}

    \item \textbf{General Recall and Precision:} Applied in the full RAG configuration. As described in Section~\ref{subsec:rag_baseline}, the LLM acts as a filter by citing only the sources actually used in the answer. Therefore, for these metrics, the set of \texttt{retrieved\_docs} consists exclusively of the chunks referenced in the final generation, making the number of retrieved documents variable:
        \[
        \text{Recall} = \frac{|\{doc \in \text{retrieved\_docs} : doc \in \text{relevant\_docs}\}|}{|\text{relevant\_docs}|}
        \]
        
        \[
        \text{Precision} = \frac{|\{doc \in \text{retrieved\_docs} : doc \in \text{relevant\_docs}\}|}{|\text{retrieved\_docs}|}
        \]

    \item \textit{Subjective metrics:} Completeness, relevance, and clarity, each evaluated by \texttt{gpt-4o} on a 1–10 scale for the generated answers obtained by concatenating the top-3 retrieved chunks. 
    In this setup, the evaluator LLM is provided only with the retrieved chunks and does not have access to the generated answers. This design ensures that the evaluation focuses on the informational quality of the retrieved passages and avoids potential feedback loops or self-evaluation bias.
    Although LLM-based evaluation cannot fully replace human judgment, it provides a scalable and replicable proxy for consistently estimating answer quality across the benchmark.
\end{itemize}

\subsection{Evaluation methods}

Three complementary evaluation methods were considered to comprehensively compare the different variants of the retrieval pipeline:

\begin{enumerate}
    \item \textit{Method 1: Classical evaluation on dense retrieval.}  
    The metrics EM, Recall@k, Precision@k, MRR\_gold, and MRR\_rel\_docs were computed directly on the retrieval system, without involving the LLM generation step.  
    This method was applied to both the custom benchmark and the \textit{SQuAD-it} dataset.

    \item \textit{Method 2: Subjective evaluation through LLM-based scoring.}
    To assess the informational quality of the retrieval, the top-$K$ retrieved chunks ($K=3$) were directly concatenated into a single text block. A separate \texttt{gpt-4o} instance then acted as an external evaluator, scoring these raw human transcripts on a 1--10 scale for completeness, relevance, and clarity. Crucially, because the evaluated text is not LLM-generated, this setup is inherently immune to stylistic ``self-preference bias''. Because the evaluated text consists exclusively of concatenated retrieved passages rather than generated answers, the evaluator LLM does not assess its own outputs. This prevents feedback loops and mitigates the well-known self-preference bias observed in LLM-as-a-judge evaluations. The exact prompt is available on GitHub \footnote{\url{https://github.com/Granataaa/educational-rag-el/blob/main/rag_api/testBench2.py}}.

    \item \textit{Method 3: Classical evaluation on the full RAG pipeline.}  
    In this case, the query was processed by the complete RAG system, where the LLM filters and synthesizes the retrieved documents.  
    The computed metrics included EM, Recall, Precision, MRR\_gold, and MRR\_rel\_docs.  
    Unlike the dense retrieval case, @k-based metrics were not used, since the number of documents returned by the LLM is not fixed.
\end{enumerate}

This evaluation framework provides the basis for the experimental analysis presented in the next section.

\section{Results}\label{sec:results}

This section presents the results obtained from the evaluation of the different pipelines. We compare the \textit{Standard RAG} baseline against the variants incorporating Entity Linking, specifically focusing on the performance of our proposed \textit{ELERAG} method relative to the alternative re-ranking strategies defined in Section~\ref{sec:methodology}:

\begin{itemize}
    \item \textit{ELERAG (RRF-Based Re-ranking)}: Our proposed method, applied to the Top-30 candidate chunks retrieved by FAISS.
    \item \textit{Hybrid Sparse-Dense Baseline (Dense + BM25)}: This hybrid approach is also applied to the Top-30 candidate chunks retrieved by FAISS. Note that this baseline was introduced specifically to isolate pure retrieval performance and was exclusively evaluated under Method 1 using our custom educational dataset.
    \item \textit{Weighted-Score Re-ranking}: Implemented with a weighting factor $\beta = 0.5$, applied to the Top-30 candidate chunks retrieved by FAISS.
    \item \textit{RRF + Cross-Encoder Re-ranking}: This method is applied to re-score the Top-20 candidates selected via RRF from an extended initial pool of 50 chunks retrieved by FAISS.
    \item \textit{Cross-Encoder (Standalone)}: Re-scoring of Top-30 candidates retrieved by FAISS.
\end{itemize}

The quantitative and qualitative findings are reported according to the three evaluation methods described in the previous section.

\subsection{Method 1: Classical Evaluation on Retrieval}

Table~\ref{tab:results_dense} summarizes the performance of the retrieval pipelines on the custom educational benchmark. To facilitate comparison, the best result for each metric is highlighted in bold.

\begin{table}[H]
\centering
\caption{Retrieval performance on the custom benchmark. The proposed \textit{RRF-Based Re-ranking} strategy achieves the best results in Exact Match (EM) and Precision@1, surpassing both the baseline and the Cross-Encoder.}
\resizebox{\textwidth}{!}{
\begin{tabular}{lcccccccc}
\toprule
\textbf{Pipeline} & \textbf{EM} & \textbf{R@1} & \textbf{R@3} & \textbf{R@5} & \textbf{R@10} & \textbf{P@1} & \textbf{MRR\_G} & \textbf{MRR\_RD}\\
\midrule
Baseline & 0.522 & 0.377 & 0.640 & 0.729 & 0.807 & 0.652 & 0.652 & 0.759\\
Sparse-Dense & 0.550 & 0.384 & \textbf{0.655} & \textbf{0.749} & 0.787 & 0.638 & 0.665 & 0.752\\
Weighted-Score & 0.536 & 0.391 & 0.647 & 0.717 & 0.795 & 0.681 & 0.654 & 0.772\\
\textbf{ELERAG (RRF)} & \textbf{0.565} & 0.399 & 0.647 & 0.725 & 0.795 & \textbf{0.696} & \textbf{0.668} & \textbf{0.779}\\
RRF+Cross-Encoder & 0.536 & \textbf{0.408} & 0.645 & 0.698 & 0.802 & 0.652 & 0.647 & 0.760\\
Cross-Encoder & 0.536 & \textbf{0.408} & 0.633 & 0.693 & \textbf{0.816} & 0.652 & 0.646 & 0.755\\
\bottomrule
\end{tabular}
}
\label{tab:results_dense}
\end{table}

Our proposed \textit{ELERAG} strategy emerged as the most effective method for this domain, achieving the highest scores in key precision metrics (EM, MRR, P@1).
This superiority demonstrates that fusing the entity signal with dense retrieval via RRF is more robust than a simple linear combination (Weighted-Score), effectively filtering out semantically similar but factually incorrect chunks.

The \textit{Standalone Cross-Encoder} achieved the highest Recall@10 ($0.816$), confirming the ability of deep transformer models to identify relevant content within a broader window. However, this broad recall did not translate into superior ranking precision: both Cross-Encoder variants scored lower than \textit{ELERAG} in EM and MRR (both MRR\_gold and MRR\_rel\_docs).
This might depend on the fact that while the generic, pre-trained Cross-Encoder captures general semantic relevance effectively, it lacks the specific domain grounding provided by the explicit Entity Linking signal. In this specialized educational context, the Cross-Encoder tends to assign high scores to semantically plausible but factually distinct chunks, whereas \textit{ELERAG} successfully utilizes Wikidata IDs to disambiguate and prioritize the exact target concept.

It is worth noting that the \textit{Hybrid Sparse-Dense Baseline} and the \textit{Standalone Cross-Encoder} maintain higher Recall@5 and Recall@10 scores, respectively, compared to \textit{ELERAG}. This might be due to the fact that while entity linking can retrieve additional relevant passages, it may also introduce some noise, as passages discussing related entities or topics may be retrieved even when they are not directly relevant to the query. The RRF step mitigates this effect by down-weighting passages that are not consistently ranked highly across signals. However, when evaluating Recall@k with high values of k, some of these additional candidates may occupy relevant positions, which can slightly reduce recall compared to other baselines. This highlights a characteristic behavior of the RRF algorithm when applied to entity-based signals: it aggressively optimizes the very top positions (Top-1) to maximize precision for the limited context window of the LLM. Although this strict filtering may push some marginally relevant documents further down the list, the substantial gain in MRR and Exact Match is far more valuable for a RAG system, where the generator's performance depends primarily on the quality of the very first retrieved chunks.

Furthermore, this comparison perfectly validates the architectural choice of integrating Entity Linking over traditional sparse retrieval. While the \textit{Hybrid Sparse-Dense Baseline} (Dense+BM25) performs competitively in those mid-tier recall metrics, \textit{ELERAG} significantly outperforms it in the most critical ranking metrics, specifically Precision@1 ($0.695$ vs. $0.637$) and both MRR metrics. This empirical evidence demonstrates that our Entity Linking module successfully goes beyond mere syntactic overlap: by grounding terms to Wikidata, it resolves disambiguations and captures domain-specific relationships that push the precise gold chunk to the very top of the ranking, which is an essential requirement for the subsequent generation phase.

\begin{figure}[htbp]
    \centering
    \includegraphics[width=0.7\linewidth]{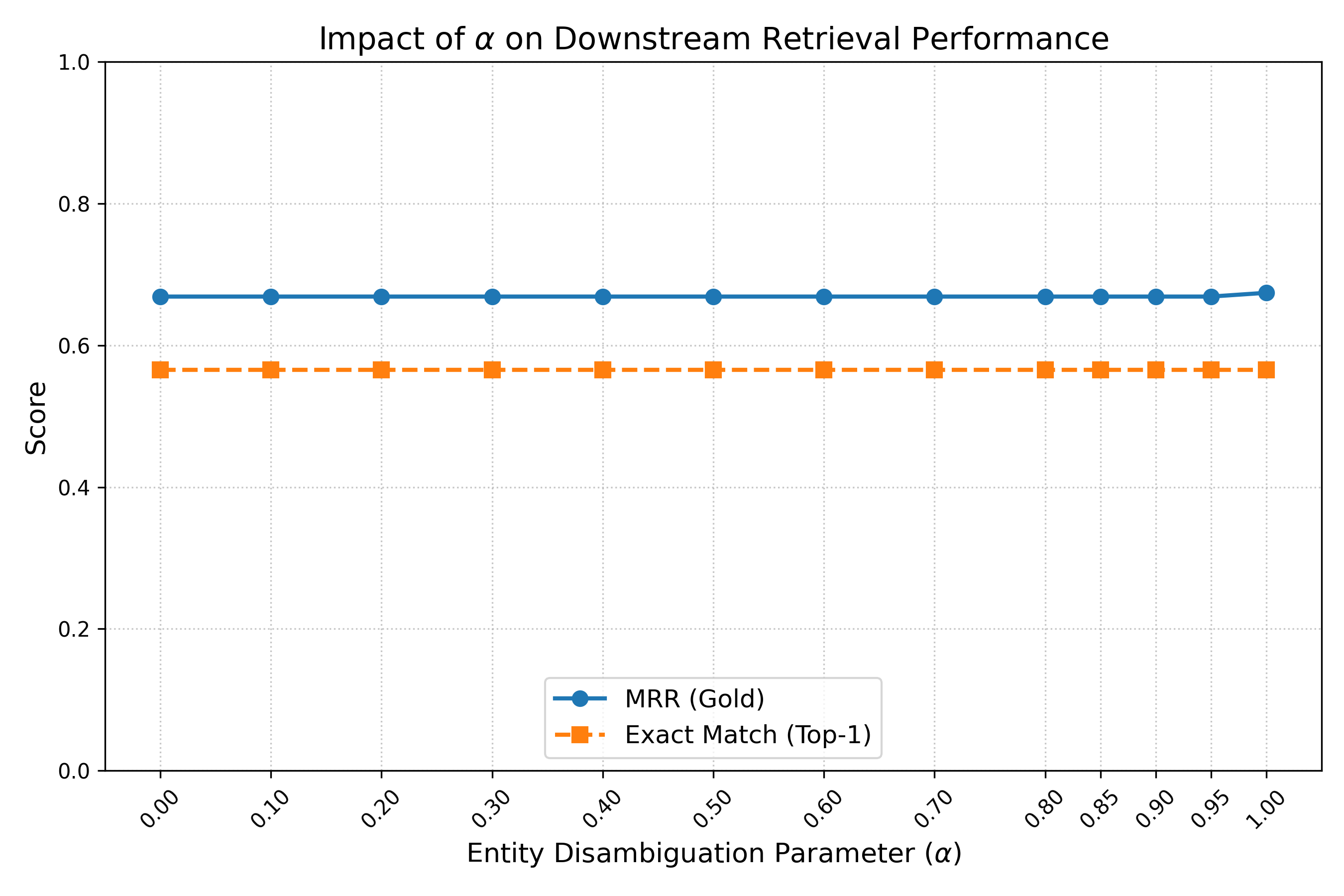}
    \caption{Ablation Study of $\alpha$ parameter in the Entity Disambiguation module. The plot shows that between 0.0 and 1.0 there are almost no variation of EM and MRR.}
    \label{fig:alpha_sensitivity}
\end{figure}

To assess the dependence on the specific entity linking model, we also conducted experiments by varying the parameter $\alpha$ (see Sect. \ref{subsec:ELmodule}). Note that when $\alpha = 0$, the ranking is based only on the Wikidata API score (i.e., popularity), while when $\alpha = 1.0$, the ranking relies exclusively on the similarity score. The results are reported in Figure \ref{fig:alpha_sensitivity}.


We observed almost no variation when $\alpha$ ranges between 0.0 and 1.0 (EM consistently equals 0.565, while MRR ranges between 0.668 and 0.674), confirming that our method is largely independent of the specific entity linking strategy employed.

\newpage
\subsection{Method 2: Subjective Evaluation through LLM-based Scoring}

Table~\ref{tab:results_qualitative} reports the average scores assigned by the evaluator LLM to the answers generated by each pipeline. The evaluation focuses on three dimensions: Completeness, Relevance, and Clarity.

\begin{table}[H]
\centering
\caption{Subjective evaluation scores (scale 1--10). \textit{ELERAG} achieves the highest scores in all categories, confirming that entity-aware retrieval leads to more comprehensive and relevant answers.}
\resizebox{0.8\textwidth}{!}{
\begin{tabular}{lccc}
\toprule
\textbf{Pipeline} & \textbf{Completeness} & \textbf{Relevance} & \textbf{Clarity} \\
\midrule
Baseline & 5.99 & 5.45 & 4.51 \\
Weighted-Score & 6.00 & 5.43 & 4.42 \\
\textbf{ELERAG (RRF)} & \textbf{6.10} & \textbf{5.57} & \textbf{4.54} \\
RRF+Cross-Encoder & 5.94 & 5.39 & 4.43 \\
Cross-Encoder & 5.83 & 5.30 & 4.32 \\
\bottomrule
\end{tabular}
}
\label{tab:results_qualitative}
\end{table}

Consistent with the quantitative metrics, the \textit{ELERAG (RRF)} strategy achieved the highest scores across all qualitative dimensions. The improvement is particularly substantial in \textit{Completeness} ($6.10$) and \textit{Relevance} ($5.57$). This indicates that balancing semantic similarity with entity popularity provides the generator with a richer and more pertinent context, allowing it to construct answers that are not only factually correct but also exhaustive.

The qualitative evaluation also highlights the limitations of general-purpose neural re-rankers in this specialized domain. The \textit{Standalone Cross-Encoder} recorded the lowest scores for Completeness and Relevance. This suggests that while the Cross-Encoder identifies semantically related content, it prioritizes chunks that are tangentially relevant but lack the specific factual definitions required for educational quality—effectively introducing "semantic noise" that the robust RRF fusion successfully filters out.

The scores for \textit{Clarity} are relatively stable across all systems (ranging from $4.42$ to $4.54$). Since the evaluation target in this method is formed by directly concatenating the raw retrieved chunks, this stability is expected: the source material (lecture transcripts) is identical for all pipelines. However, the slight edge held by \textit{ELERAG} ($4.54$) suggests that the entity-driven retrieval tends to select more structurally complete or self-contained passages, resulting in a slightly more coherent reading flow compared to the other methods.

\subsection{Method 3: Classical Evaluation on the Full RAG Pipeline}

Table~\ref{tab:results_rag} presents the performance metrics calculated on the final output of the end-to-end RAG system. In this evaluation, the metrics consider only the chunks explicitly cited by the LLM in the generated answer, effectively treating the generator as a final relevance filter.

\begin{table}[H]
\centering
\caption{Performance of the full RAG system (post-LLM filtering). \textit{ELERAG} maintains its superiority, achieving the highest Exact Match and MRR, proving that better initial ranking translates to better generation support.}
\resizebox{\textwidth}{!}{
\begin{tabular}{lccccc}
\toprule
\textbf{Pipeline} & \textbf{EM} & \textbf{Recall} & \textbf{Precision} & \textbf{MRR\_G} & \textbf{MRR\_RD} \\
\midrule
Baseline & 0.522 & 0.577 & 0.428 & 0.603 & 0.714 \\
Weighted-Score & 0.522 & 0.563 & 0.448 & 0.599 & 0.729 \\
\textbf{ELERAG (RRF)} & \textbf{0.551} & \textbf{0.589} & \textbf{0.458} & \textbf{0.622} & \textbf{0.742} \\
RRF+Cross-Encoder & 0.507 & 0.582 & 0.441 & 0.589 & 0.708 \\
Cross-Encoder & 0.522 & \textbf{0.589} & 0.437 & 0.599 & 0.713 \\
\bottomrule
\end{tabular}
}
\label{tab:results_rag}
\end{table}

The results confirm that the retrieval quality improvements propagate effectively to the final RAG output. \textit{ELERAG (RRF)} consistently outperforms all other configurations, achieving the highest scores in Exact Match ($0.551$) and MRR ($0.622$). Notably, while it ties with the \textit{Standalone Cross-Encoder} for the highest Recall ($0.589$), it converts this into significantly better Precision ($0.458$ vs $0.437$). This demonstrates that the Entity-Aware RRF strategy provides a cleaner context, enabling the LLM to identify and cite the correct gold answer more frequently.

The comparison with the \textit{Standalone Cross-Encoder} further reinforces the importance of ranking precision. Despite matching ELERAG in retrieval recall, the Cross-Encoder performs identically to the Baseline in Exact Match ($0.522$). This discrepancy suggests that the neural model successfully retrieves relevant documents but fails to rank the specific gold chunk effectively, leading the LLM to synthesize information from a mix of relevant documents rather than focusing on the single best source. The \textit{RRF+Cross-Encoder} pipeline showed a slight degradation (EM $0.507$), confirming that increasing complexity does not yield additive benefits in this setup.

\subsection{Evaluation on a standard benchmark (SQuAD-it)}

Table~\ref{tab:results_squad} reports the retrieval performance on the SQuAD-it dataset. This benchmark serves as a control experiment on a general-domain corpus (Wikipedia).

\begin{table}[H]
\centering
\caption{Retrieval performance on SQuAD-it (General Domain). Unlike the educational dataset, here the \textit{Cross-Encoder} achieves the best results, highlighting its strength on standard web-like data.}
\resizebox{\textwidth}{!}{
\begin{tabular}{lccccccc}
\toprule
\textbf{Pipeline} & \textbf{EM} & \textbf{R@1} & \textbf{R@3} & \textbf{R@5} & \textbf{R@10} & \textbf{P@1} & \textbf{MRR} \\
\midrule
Standard RAG (Baseline) & 0.693 & 0.693 & 0.843 & 0.884 & 0.922 & 0.693 & 0.776 \\
Weighted-Score & 0.645 & 0.645 & 0.804 & 0.853 & 0.904 & 0.645 & 0.735 \\
ELERAG (RRF) & 0.672 & 0.672 & 0.829 & 0.875 & 0.922 & 0.672 & 0.760 \\
\textbf{RRF+Cross-Encoder} & \textbf{0.777} & \textbf{0.777} & \textbf{0.885} & \textbf{0.912} & 0.936 & \textbf{0.777} & \textbf{0.836} \\
\textbf{Cross-Encoder} & 0.776 & 0.776 & \textbf{0.885} & \textbf{0.912} & \textbf{0.938} & 0.776 & \textbf{0.836} \\

\bottomrule
\end{tabular}
}
\label{tab:results_squad}
\end{table}

On the SQuAD-it dataset, the performance trend is sharply reversed. The \textit{Cross-Encoder} configurations achieved the best results across all metrics (EM $\approx 0.777$, MRR $0.836$), significantly outperforming both the baseline and the entity-based methods (\textit{ELERAG} EM $0.672$). This can be explained by the fact that this dataset represents a highly in-domain scenario. SQuAD-it is derived from Wikipedia, which closely resembles the type of data commonly used to train or pre-train many embedding models and cross-encoders. In such settings dense and cross-encoder models tend to perform particularly well because the lexical, stylistic, and semantic characteristics of the queries and passages closely match their training distribution. This suggests that in general-domain texts like Wikipedia, which are linguistically standard and factually dense, the semantic signal of pre-trained models is robust enough, and the additional entity enforcement does not provide the disambiguation benefit observed in the lecture corpus.

A marginal trade-off is observed between the Cross-Encoder variants: while the \textit{RRF + Cross-Encoder} pipeline achieves a slight advantage in precision metrics, the \textit{Standalone} version retains a higher Recall@10 ($0.938$). This indicates that RRF acts as a pre-filter, refining the candidate pool for precision at the cost of slightly reducing deep recall.

This divergence provides strong experimental evidence for the \textit{Domain Mismatch} hypothesis. The Cross-Encoder leverages its pre-training on general web data to rank standard text effectively. However, as demonstrated in the previous sections, this advantage disappears in the high-ambiguity domain of university lectures. This leads to a crucial conclusion: while Cross-Encoders are optimal for general tasks, they are not a universal solution. For specialized educational platforms where training data is scarce, our entity-aware approach provides a more robust retrieval mechanism without the computational cost of a heavy neural re-ranker.

\subsection{Discussion}\label{sec:discussion}

The experimental results present a clear and divergent pattern between the standard SQuAD-it benchmark and the  specific educational dataset, which forms a major finding of this study.

First, the quantitative analysis on our specialized educational corpus demonstrates that our proposed \textit{ELERAG} architecture achieved the best overall performance. It consistently outperformed not only the \textit{Weighted-Score} baseline but also the \textit{Standalone Cross-Encoder}.
Crucially, while the Standalone Cross-Encoder showed high retrieval capacity (matching ELERAG in Recall), it failed to achieve comparable precision in Exact Match and MRR. This indicates that in a specialized domain with subtle terminological nuances, a generic deep-learning re-ranker may struggle to prioritize the specific "gold" chunk among other semantically plausible candidates. \textit{ELERAG}, by leveraging the explicit entity signal, successfully bridges this gap.

Conversely, on the general-purpose \textit{SQuAD-it} benchmark, this trend inverted sharply: the \textit{Cross-Encoder} configurations obtained the highest scores across all metrics, proving their superiority in a standard, general-domain QA setting.
This quantitative divergence provides strong experimental evidence for our \textit{Domain Mismatch} hypothesis. It indicates that while large, pre-trained re-rankers excel on web-style data (like Wikipedia), a domain-adapted hybrid model like \textit{ELERAG} is significantly more effective on specialized, narrative corpora (like university lectures). In these contexts, the explicit signals provided by Entity Linking align better with the data than the generic semantic patterns learned by the Cross-Encoder.

Given that \textit{ELERAG} appears to be the most effective solution for the specific domain, qualitative analysis helps to explain why the integration of Entity Linking provides this advantage.
Inspection of the results confirms that the entity-based signal is crucial in high-ambiguity scenarios. A clear example is the query ``Who is Smith?''. The \textit{Baseline} system, relying only on semantic similarity, retrieved scattered and less relevant documents based on broad keyword matching. In contrast, the \textit{ELERAG} pipeline, guided by the unambiguous entity ID from Wikidata, successfully promoted the correct, relevant passages to the top ranks. This allowed the LLM to generate a more coherent, factually dense, and precise answer. This qualitative benefit was less pronounced in broad, low-ambiguity queries, where the baseline's semantic search was already sufficient. This suggests that the primary value of our hybrid approach lies in its ability to resolve factual ambiguity, which is a critical weakness of purely semantic systems in technical domains. Moreover, because the system relies on multilingual embeddings and Wikidata identifiers shared across languages, the overall architecture is largely language-agnostic and could be applied to other languages with comparable knowledge base coverage.

Beyond pure retrieval performance, it is crucial to highlight the significant difference in computational cost at inference time. The \textit{Standalone Cross-Encoder} relies on pairwise evaluations: to re-rank $N$ retrieved chunks, it requires $N$ distinct forward passes through a deep Transformer model analyzing the concatenated query-document strings. Due to the $\mathcal{O}(L^2)$ complexity of the self-attention mechanism (where $L$ is the sequence token length), this process is computationally expensive and typically requires dedicated GPU hardware to achieve low latency in real-time RAG applications. Conversely, \textit{ELERAG} shifts the heavy computational burden to the offline indexing phase, where chunk entities are pre-extracted. At query time, the system only processes the short user query through a lightweight NER module, performs standard I/O API calls to Wikidata for the extracted mentions, and computes the final ranking via a simple Boolean set intersection of Wikidata IDs. This highly efficient, CPU-friendly operation avoids heavy neural inference, making it exceptionally suited for scalable, domain-specific deployments.

Another practical consideration concerns the use of automatically transcribed lecture data. Speech recognition may occasionally introduce spelling variations or incomplete entity mentions. While such cases may reduce the effectiveness of the entity linking signal, the hybrid architecture of ELERAG ensures robustness by combining entity-aware retrieval with dense semantic retrieval. This design prevents isolated transcription errors from significantly degrading the overall system performance.

Finally, the analysis confirmed the robustness of the full RAG pipeline in handling out-of-domain questions. Thanks to the LLM acting as a semantic filter, irrelevant queries (e.g., ``What is the capital of France?'') correctly resulted in a safe fallback response (``No relevant information found''), effectively mitigating model hallucination and reinforcing the system's reliability as a tutoring tool.

\section{Conclusion}

We presented \textit{ELERAG}, an enhanced RAG architecture that integrates a factual signal derived from entity linking to improve the accuracy of educational question-answering systems in Italian. Our system has the potential to enhance RAG systems in domains that differ from standard RAG benchmarks and training data, such as the educational domain.

Our experimental evaluation offers several important implications for research in retrieval-augmented generation and educational AI. First, the consistent improvements obtained through entity-based re-ranking demonstrate that factual grounding can be enhanced without the need for expensive model retraining, solely through structured post-retrieval refinement. Crucially, our experiments demonstrated that injecting structured semantic knowledge (Wikidata) yields superior top-ranking performance (Precision@1 and MRR) compared to strong traditional exact-match baselines like BM25, proving that semantic disambiguation is highly beneficial in specialized educational domains. Second, the contrast between ELERAG and Cross-Encoder performance emphasizes that high-capacity models do not generalize uniformly across all domains: their success strongly depends on the alignment between pre-training data and the target corpus. This finding suggests that lightweight hybrid methods—such as our entity-enriched retrieval—can be a more efficient and accurate alternative to SOTA re-rankers in domain-specific applications, particularly in low-resource or multilingual settings such as Italian educational materials. Finally, the integration of an entity linking module provides a clear framework for combining symbolic and neural representations, showing how explicit knowledge bases like Wikidata can complement dense embeddings to improve interpretability and precision.




Further research directions include exploring adaptive weighting schemes for the RRF, dynamically adjusting the balance between semantic and entity signals based on the query type (e.g., factual vs. conceptual). Moreover, expanding the evaluation to multilingual data and incorporating human-in-the-loop assessments would provide a richer understanding of the system's real-world educational value.

\vspace{6pt} 





\authorcontributions{Conceptualization, F.G., F.P. and M.M.; methodology, F.G., F.P. and M.M.; software, F.G.; validation, F.G., F.P. and M.M.; data curation, F.G. and F.P.; writing---original draft preparation, F.G.; writing---review and editing, F.G., F.P.. and M.M.; supervision, F.P. and M.M.; funding acquisition, F.P.. All authors have read and agreed to the published version of the manuscript.}

\funding{We acknowledge financial support from the
PNRR project Learning for All (L4ALL) funded by
the Italian MIMIT (number: F/310072/01-05/X56).}

\institutionalreview{Not applicable.}

\informedconsent{Not applicable.}

\dataavailability{The source code for the system is publicly available on GitHub at \url{https://github.com/Granataaa/educational-rag-el}. The custom dataset generated and analyzed during this study is not publicly available because it has been built from proprietary data from a private company.}

\acknowledgments{During the preparation of this study, the author(s) used GPT-4o by OpenAI for the purposes of generation and validation of the benchmark and formatting the response of the system. The authors have reviewed and edited the output and take full responsibility for the content of this publication.}

\conflictsofinterest{The authors declare no conflicts of interest.} 



\abbreviations{Abbreviations}{
The following abbreviations are used in this manuscript:
\\

\noindent 

\begin{tabular}{@{}ll}
RAG & Retrieval-Augmented Generation \\
LLM & Large Language Models \\
EL & Entity Linking \\
RRF & Reciprocal Rank Fusion \\
EM & Exact Match \\
MRR & Mean Reciprocal Rank \\
NER & Named Entity Recognition \\
EAI & Educational Artificial Intelligence \\
FAISS & Facebook AI Similarity Search \\
GPT-4o & Generative Pre-trained Transformer 4.0 \\
\end{tabular}
}

\begin{adjustwidth}{-\extralength}{0cm}

\reftitle{References}


\bibliography{bibliography.bib}

\PublishersNote{}
\end{adjustwidth}
\end{document}